\documentclass[
reprint,
 superscriptaddress,
 amsmath,amssymb,
 aps,
]{revtex4-2}

\usepackage{graphicx}
\usepackage{caption}
\usepackage{overpic}
\usepackage{subcaption}
\usepackage{dcolumn}
\usepackage{bm}
\usepackage{color}
\usepackage{xurl}
\usepackage[hidelinks]{hyperref}
\hypersetup{pdfborder={0 0 0}, breaklinks=true, pdfhighlight=/N}
\usepackage{microtype}
\emergencystretch{1em}

\usepackage[mathlines]{lineno}

\begin{document}


\title{Many-body activity emerging in a monolayer of air-fluidized hollow pentagons}

\author{Wan-Jung Lin}
\affiliation{Department of Mechanical Engineering, University of Michigan, Ann Arbor.}
\author{Bohan Wu-Zhang}
\affiliation{Departamento de Estructura de la Materia, Física Térmica y Electrónica, Universidad Complutense de Madrid, Madrid, Spain.}
\affiliation{GISC - Grupo Interdiplinar de Sistemas Complejos, Madrid, Spain.}
\author{Rodrigo Fernández-Quevedo García}
\affiliation{Departamento de Estructura de la Materia, Física Térmica y Electrónica, Universidad Complutense de Madrid, Madrid, Spain.}
\affiliation{GISC - Grupo Interdiplinar de Sistemas Complejos, Madrid, Spain.}
\author{C. Miguel Barriuso G.}
\affiliation{Departamento de Estructura de la Materia, Física Térmica y Electrónica, Universidad Complutense de Madrid, Madrid, Spain.}
\affiliation{GISC - Grupo Interdiplinar de Sistemas Complejos, Madrid, Spain.}
\author{Aayush Agarwal}
\affiliation{Department of Mechanical Engineering, University of Michigan, Ann Arbor.}
\author{Yue Fan}
\affiliation{Department of Mechanical Engineering, University of Michigan, Ann Arbor.}
\author{Miguel Ruiz-García}
\affiliation{Departamento de Estructura de la Materia, Física Térmica y Electrónica, Universidad Complutense de Madrid, Madrid, Spain.}
\affiliation{GISC - Grupo Interdiplinar de Sistemas Complejos, Madrid, Spain.}
\author{Chantal Valeriani}
\affiliation{Departamento de Estructura de la Materia, Física Térmica y Electrónica, Universidad Complutense de Madrid, Madrid, Spain.}
\affiliation{GISC - Grupo Interdiplinar de Sistemas Complejos, Madrid, Spain.}
\author{Hongyi Xiao}
\affiliation{Department of Mechanical Engineering, University of Michigan, Ann Arbor.}



\date{\today}

\begin{abstract}

Particles governed by many-body interactions exhibit remarkably complex structures and dynamics. We experimentally investigate a monolayer of pentagon particles subjected to an up-lifting air flow which induces many-body aerodynamic interactions and stochastic motion akin to a thermal bath. To minimize air flow resistance, particles move collectively with interactions dictated by their geometry: hollow particles exhibit effective attraction, whereas solid particles repel each other. Under sufficiently large air flow, sparsely packed hollow pentagons overcome substrate friction and undergo long-time diffusive motion. Under lower air flow, we see a coexistence of isolated, static pentagons and densely packed, ``active'' clusters, whose particles display super-diffusivity. This ``emergent activity'' arises collectively when locally disordered structures interact with the air flow, resulting in correlated motion across broad temporal and spatial scales. Using Langevin dynamics simulations of two-dimensional attractive active pentagons, whose activity is an effective result of the local packing density, we further unravel the basic features of this emergent activity.

\end{abstract}

\maketitle

\section{\label{sec:level1} Introduction}

The celebrated saying “More is different” by Anderson~\cite{Anderson1972} is frequently invoked to describe the emergent phenomena observed in systems of simple particles such as hard spheres~\cite{torquato2010jammed} and Lennard-Jones particles~\cite{frenkel1980computer,debenedetti2001supercooled}, even though the underlying interaction potentials themselves remain unchanged as the number of neighbors increases.
For many real-world systems with a complex nature, however, both individual particle's dynamics and interaction potentials can be influenced by the density, position, and orientation of multiple neighboring particles. 
Such many-body interactions can be found in a variety of living systems such as bacteria~\cite{zhang2010collective,Beer2019,be2020phase}, social animals~\cite{Zampetaki2021,Zampetaki2024}, and human crowds~\cite{wirth2023neighborhood,gu2025emergence}. 
Many-body interactions are also important in physical systems such as colloidal suspensions in the presence of electrical charges, polymers, and nano-particles~\cite{dobnikar2002many,dijkstra2006effect,parolini2015interaction} and metallic glasses with ions among electron clouds~\cite{cheng2009atomic,Li2024}. Understanding how many-body effects influence the collective behavior of such particle systems can enable us to use more complicated building blocks to create advanced condensed matter systems.

Many-body effects introduce structural and dynamical features absent under purely pairwise interactions, even though a complete description of their influence is still lacking.
Living systems such as social animals and bacteria often exhibit collective behaviors across large length and time scales for optimizing a shared field such as temperature and resource concentration~\cite{mckechnie2001thermoregulation,ocko2014collective,Zampetaki2021}, giving the corresponding thermo- and chemo-taxis of individuals a many-body nature.
When approximated by a particle system with three-body interactions, the resulting packing structures in clusters display lower periodicity than those organized via pairwise interactions, while allowing a larger portion of particles to reach the optimal field state~\cite{Zampetaki2021}. 
Collective motion, such as flocking, schooling, and swarming, can also have many-body effects~\cite{tevrugt2025}. An individual bacterium can remain static when isolated, while a dense cluster may be highly mobile collectively, which is also coupled to their spatial arrangement as bacteria have anisotropic shapes~\cite{Beer2019,be2020phase}. 
However, for living matter, the observed collective behaviors can have physiological and biochemical origins that can be difficult to decouple from physical mechanisms in experiments. 

For inanimate matter, it is possible to experimentally design particle systems to amplify many-body effects. Quincke rollers, for example, are dielectric colloids that can spontaneously rotate and self-propel due to an electrohydrodynamic instability when experiencing a DC electric field exceeding a certain threshold~\cite{bricard2015emergent,brosseau2017electrohydrodynamic,das2019active}. However, this threshold may be lowered if particles are in close proximity to each other, inducing many-body effects via distorting the local electric field and coupled hydrodynamics\cite{Bricard2013, Liu2021}. 
Many-body effects also exist in certain particles that are either suspended or at liquid interfaces~\cite{dalbe2011aggregation,kokot2018manipulation,han2020emergence,xiao2020strain,xiao2023identifying,yan2026densely,hobson2026structural}, although their implications have not been explicitly discussed. 
Different from the heavily studied wet and highly dissipative systems, airborne and inertial particles can also be configured to have many-body interactions, such as in dusty plasma~\cite{yu2025physics} and in acoustic levitation~\cite{lim2022mechanical,brown2024direct,lim2024acoustic}. In such systems, many-body interactions can trigger emergent collective transport via multiple-scattering forces in acoustic fields~\cite{lim2024acoustic} and via non-reciprocal wake forces in dusty plasmas, allowing static grains to transition into mobile, fluid-like states~\cite{yu2025physics}. In these cases, the interaction field acts as a conduit for energy transfer that an isolated particle could not exploit.

\begin{figure}[b!]
\centering
\includegraphics[width=0.97\linewidth]{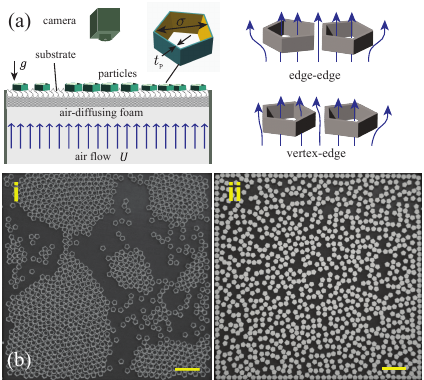}
\vspace{-2 pt}
\caption{(a) Experimental schematic (not to scale) and (b) snapshots of pentagons experiencing regularized up-lifting air flow. The particles have an altitude of $\sigma=7.2$\,mm and a height of 3\,mm. Panel i of (b) shows hollow particles with a wall thickness $t_\mathrm{p}=1.2$\,mm, and ii shows solid particles. Scale bar: 4\,cm. 
}
\label{fig:app}
\end{figure}

In this work, we investigate the influence of many-body effects on the dynamics and structure in a system of air-fluidized particles. Unlike other suspended particle systems like colloids that are mediated by quiescent fluids, here we use fast air flows to induce aerodynamic particle interactions that are many-body in nature~\cite{capecelatro2015fluid,huck2018role,mema2021fluidization}. Although in many engineering contexts, particles move along the air flow~\cite{sundaresan2000modeling,capecelatro2024gas}, we study the in-plane dynamics of a horizontal monolayer of particles under an air flow that is up-lifting while still sub-levitating~\cite{Ojha2004,abate2005partition,abate2006approach,keys2007measurement,Daniels2009,Xiao2022}, see Fig.~\ref{fig:app}a. The air flow induces stochastic lateral forces on the particles, e.g., from random aerodynamic wakes, which overcome the particles' friction with the substrate, i.e., an air-permeable screen.
Previous studies on this subject have shown that an isolated sphere behaves like a Brownian particle with its dynamics following a Langevin equation~\cite{Ojha2004}, whereas a dense sphere packing shows sub-diffusive and diffusive particle transport like a super-cooled liquid~\cite{abate2007topological,keys2007measurement}. However, the associated many-body effects have not been discussed.

We amplify many-body effects by tuning geometries of pentagon-shaped particles, see Fig.~\ref{fig:app}a. Unlike spheres, pentagons have the flat outer edges that create narrow gaps between nearby particles, significantly constricting and biasing the air flow, and thereby induce strong coupling of particle motion via the interstitial air pressure field; particles thus move to minimize their collective resistance to the air flow. This is analogous to the aforementioned theoretical work in which ``penguins'' move collectively to optimize their temperatures~\cite{Zampetaki2021}. 

We further introduce a many-body attraction between nearby particles by making them hollow, resulting in particle clustering, see Fig.~\ref{fig:app}bi. Note that similar to the previously studied spheres, solid particles repel each other and form gas-like structures (Fig.~\ref{fig:app}bii). To be more precise, the aerodynamic interaction between a pair of edge-edge aligned hollow pentagons is repulsive at short range and then becomes attractive over a distance comparable to one pentagon altitude $\sigma$, see SI. 

Finally, the pentagon shape prevents particles from perfectly tiling the 2D space~\cite{schilling2005monte,zhao2019jamming}, and the induced geometric frustration makes the coupling between local structure and collective dynamics non-trivial. A simple example is depicted in Fig.~\ref{fig:app}a, where two nearby pentagons with an edge-edge alignment perturb the air flow symmetrically, whereas a vertex-edge alignment induces asymmetric air flow, possibly resulting in net lateral forces on the particles. This mechanism is absent for spheres.

\section{Diffusive behavior of an isolated hollow pentagon}

We first unravel the particle dynamics in the dilute limit by monitoring isolated hollow pentagons of size $\sigma=7.2$\,mm under an air speed of $U=3.5$\,m/s. The corresponding Reynolds number is $\mathrm{Re}=U\sigma/\nu=1680$, where $\nu=1.5\times10^{-5}$\,m$^2$/s is the kinematic viscosity of air. This corresponds to a flow regime in which a particle sheds unsteady and irregular vortices~\cite{kundu2024fluid}, thereby itself undergoing stochastic motion.

\begin{figure}[b!]
\centering
\vspace{-10pt}
\includegraphics[width=\linewidth]{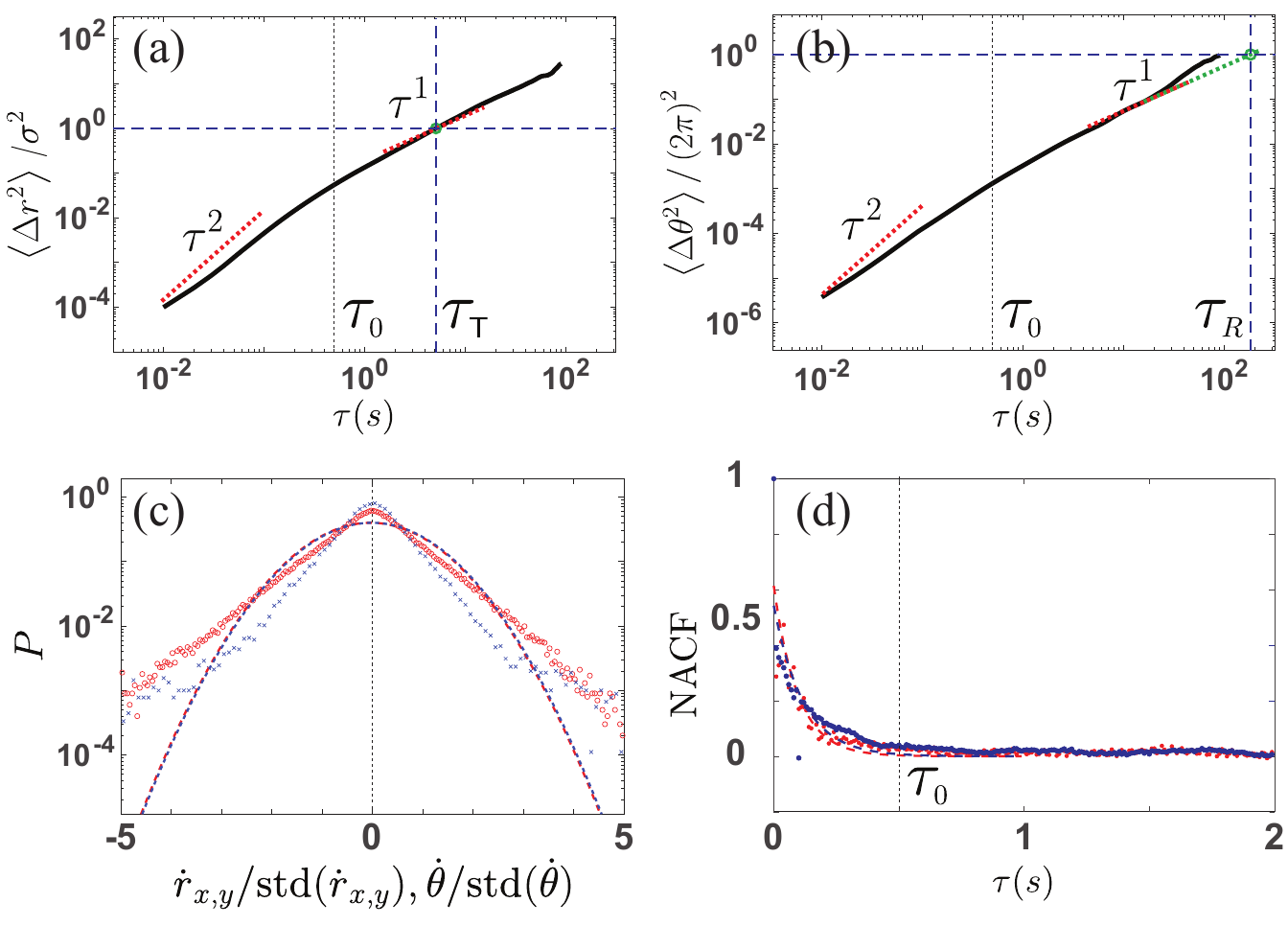}
\vspace{-2pt}
\caption{Dynamics of isolated hollow particles. (a) Translational and (b) rotational (normalized) mean squared displacement as a function of delay time, averaged over 10 randomly selected particles. In both panels, the red dashed lines indicate $\tau^2$ and $\tau^1$, while the black lines mark $\tau_0=0.5$\,s. The blue reference lines mark $\tau_{\mathrm{T}}=5.1$\,s in (a) and $\tau_{\mathrm{R}}=183.4$\,s in (b), where normalized MSD$=1$ (green circles). (c) Probability distributions of translational (red) and rotational (blue) velocities, normalized by their standard deviations, with Gaussian fits (dashed curves). (d) Normalized velocity autocorrelation of translational (red) and angular (blue) velocities; both lose memory within $\tau_0=0.5$\,s, marked by the black dotted line.
}
\label{fig:singleP}
\end{figure}

\begin{figure*}[t!]
\centering
\vspace{-5 pt}
\includegraphics[width=1.0\linewidth]{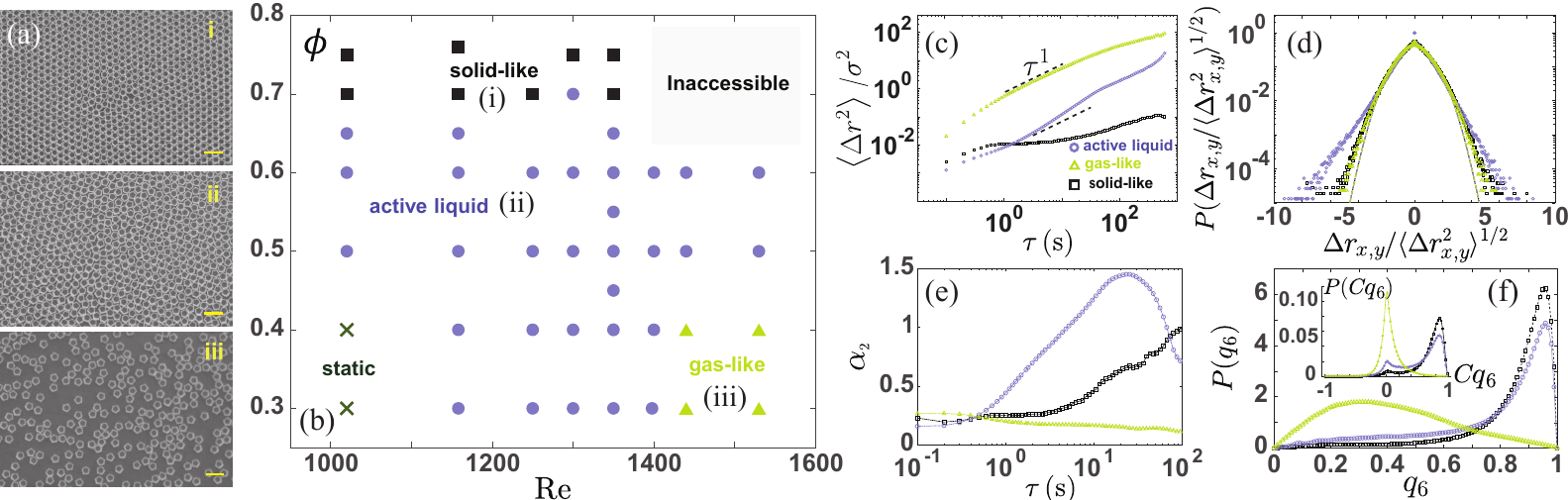}
\caption{ Phase behavior of the air-fluidized hollow pentagon system. (a) Experimental snapshots of solid-like (i), active-liquid (ii), and gas-like (iii) phases. Scale bars: 2\,cm. (b) Phase diagram as a function of Reynolds number and packing fraction, including static (crosses), solid-like (black squares), active liquid (purple circles), gas-like (green triangles), and the inaccessible region (gray). (c) Representative MSD curves for example solid-like ($\phi=0.7$, $\mathrm{Re}=1350$, black squares), active liquid ($\phi=0.5$, $\mathrm{Re}=1300$, purple circles), and gas-like ($\phi=0.3$, $\mathrm{Re}=1530$, green triangles) cases. (d) Displacement distributions of $\Delta r_{x,y}$ normalized by $\langle\Delta r_{x,y}^2\rangle^{1/2}$ over $\tau=2.5$\,s for the same three examples; dashed line is a Gaussian reference. (e) Non-Gaussian parameter $\alpha_2$ as a function of delay time $\tau$ for the three examples. (f) Distribution of the bond-orientational order parameter $q_6$ for the three examples. Inset shows the distribution of the $q_6$ correlator $C_{q_6}$. 
}
\label{fig:msd_phase}
\end{figure*}

Based on each particle's position $\vec r$ and orientation $\theta$, we calculate its mean squared displacement (MSD), $\langle\Delta r^2\rangle=\langle\left[ \vec{r}(t+\tau) - \vec{r}(t) \right]^2 \rangle$, and mean squared angular displacement, $\langle\Delta \theta^2\rangle=\langle\left[ \theta(t+\tau) - \theta(t) \right]^2 \rangle$, with $t$ being the wait time and $\tau$ the delay time. The average slopes of the translational and rotational MSDs approach one on the log-log scale at long times, see Fig.~\ref{fig:singleP}a,b, indicating long-term diffusive motion. This can be further verified by the normalized auto-correlation functions (NACFs) of translational and rotational velocities, $\left\langle \dot{\vec{r}}(t) \cdot \dot{\vec{r}}(t+\tau) \right\rangle /\left\langle \dot{\vec{r}}(t) \cdot \dot{\vec{r}}(t) \right\rangle $ and  $\left\langle {\dot{\theta}}(t) {\dot{\theta}}(t+\tau) \right\rangle /\left\langle {\dot{\theta}}(t)^2 \right\rangle $, respectively. In Fig.~\ref{fig:singleP}d, both NACFs decay to zero after a characteristic time, $\tau_0\approx0.5$\,s, indicating that a particle's motion de-correlates on this time scale, consistent with the diffusive regime observed in the MSDs. 

We then examine the translational and rotational velocity distributions in Fig.~\ref{fig:singleP}c, which show higher peaks than that of the Gaussian distribution. This is likely due to the fact that the pentagons are only sub-levitated and Coulomb friction can result in stick-slip behaviors at short times. At lower air flows, an isolated particle may become completely static due to friction. However, the particle dynamics is significantly different when more particles are nearby, as discussed next. 

\section{Phase behaviors of collective dynamics}

To understand collective dynamics of hollow pentagons, we performed a parametric sweep across various air speeds $U$ and area fractions $\phi$, which reveals distinct solid-, liquid-, and gas-like behaviors, with representative snapshots shown in Fig.~\ref{fig:msd_phase}a. A phase diagram is mapped out in Fig.~\ref{fig:msd_phase}b based on the degree of clustering and the MSD of the particles, see examples in Fig.~\ref{fig:msd_phase}c. More specifically, we fit the MSDs to $\langle\Delta r^2\rangle \sim\tau^\alpha$, and examine $\alpha$.

While the system remains static at low air speeds and low packing fractions ($\mathrm{Re} \le 1020$, $\phi \le 0.4$), increasing either $U$ or $\phi$ results in a liquid-like regime with non-zero particle motion. While this is expected for increasing $U$ (as with an isolated particle), the mobilization of particles simply due to increased $\phi$ is the first evidence of many-body effects in our system.
In this regime, particles form clusters due to the aerodynamic attraction (snapshot ii), in which they exhibit super-diffusive transport with $\alpha > 1$. Therefore we refer to this regime as ``active liquid.'' In the meantime, sparsely packed particles can coexist with the clusters (Fig.~\ref{fig:app}bi), but they are far less active.

At even higher air speed, $\mathrm{Re} \ge 1440$, yet low area fractions, $\phi \le 0.4$, the clusters break up (snapshot iii) and particles show a normal diffusive behavior with $\alpha \approx 1$, resembling that of a gas. For $\phi\ge 0.7$, particles fully occupy the domain and their movement becomes constrained like particles in a solid, as indicated by a plateaued MSD with $\alpha \ll 1$. In this case, ordered packing structures are visible in snapshot i, despite the five-fold symmetry of the pentagon particle shape. For $\phi>0.7$ and $\mathrm{Re}>1530$, particles are fully levitated and develop out-of-plane motion, which is labeled as ``inaccessible.''

To further characterize particle dynamics, we analyzed the distributions of particle displacement components, $\Delta r_{x,y}$, for representative cases of each phase, shown in Fig.~\ref{fig:msd_phase}d. 
For each distribution, we calculate the non-Gaussian parameter as a function of delay time~\cite{Martin-Roca21},
\begin{equation} 
\alpha_2(\tau)=\frac{\left\langle \Delta r^4(\tau) \right\rangle}{2\left\langle \Delta r^2 (\tau) \right\rangle ^2}-1, 
\end{equation} 
with a Gaussian distribution corresponding to $\alpha_2=0$. This roughly applies to the gas-like case with $\alpha_2$ decaying towards zero in the long term (Fig.~\ref{fig:msd_phase}e). In contrast, both the solid-like and the active-liquid cases show fat-tailed displacement distributions with $\alpha_2$ increasing with $\tau$, indicating enhanced particle movement beyond purely thermal-like fluctuations. This differs qualitatively from that of isolated particles in Fig.~\ref{fig:singleP}.

To quantify the observed structure, we computed the bond-orientational order parameter~\cite{Steinhardt1983} 
for each particle $j$ using its nearest neighbors identified by Voronoi tessellation,
\begin{equation}
q_6^j=\frac{1}{N_j}\sum_{k=1}^{N_j}e^{i6\theta_{jk}}, 
\end{equation}
where $N_j$ is the number of nearest neighbors of $j$ and $\theta_{jk}$ is the angle between a reference axis and the bond connecting particle $j$ and its neighbor $k$. The resulting distribution of $q_6$ in the gas-like case (Fig.~\ref{fig:msd_phase}f) peaks around $q_6=0.3$--$0.4$, while that in the solid-like case peaks near $q_6=0.9$, indicating strong local six-fold symmetry. The active-liquid case also shows a peak near $q_6=0.9$ but with a broader distribution than that of the solid-like case. These phases can be further distinguished by the $q_6$ correlator~\cite{vanMeel2012},
\begin{equation}
C_{q_6}(j,k)=\frac{\vec{q}_6(j)\cdot\vec{q}^*_6(k)}{|\vec{q}_6(j)|\cdot|\vec{q}_6(k)|}, 
\end{equation}
where $(j,k)$ indicates particle pairs within a given region of interest. Note that only here do we invoke the complex notation for $q_6$. The $C_{q_6}$ distribution (inset of Fig.~\ref{fig:msd_phase}f) shows a peak near zero for both the gas-like and active-liquid cases, indicating little long-range orientational order, while the solid-like case exhibits a single peak near $C_{q_6}\approx0.9$. This further shows that the active-liquid clusters have short-range order and long-range disorder. 

Finally, we note that the boundaries between different regimes are likely not sharp. E.g., the behavior of a liquid-like system at Re$=1400$ and $\phi=0.5$ would start to approximate gas-like behaviors.

\section{Many-body effects induce super-diffusivity}

\begin{figure*}[t!]
\centering
\includegraphics[width=0.99\linewidth]{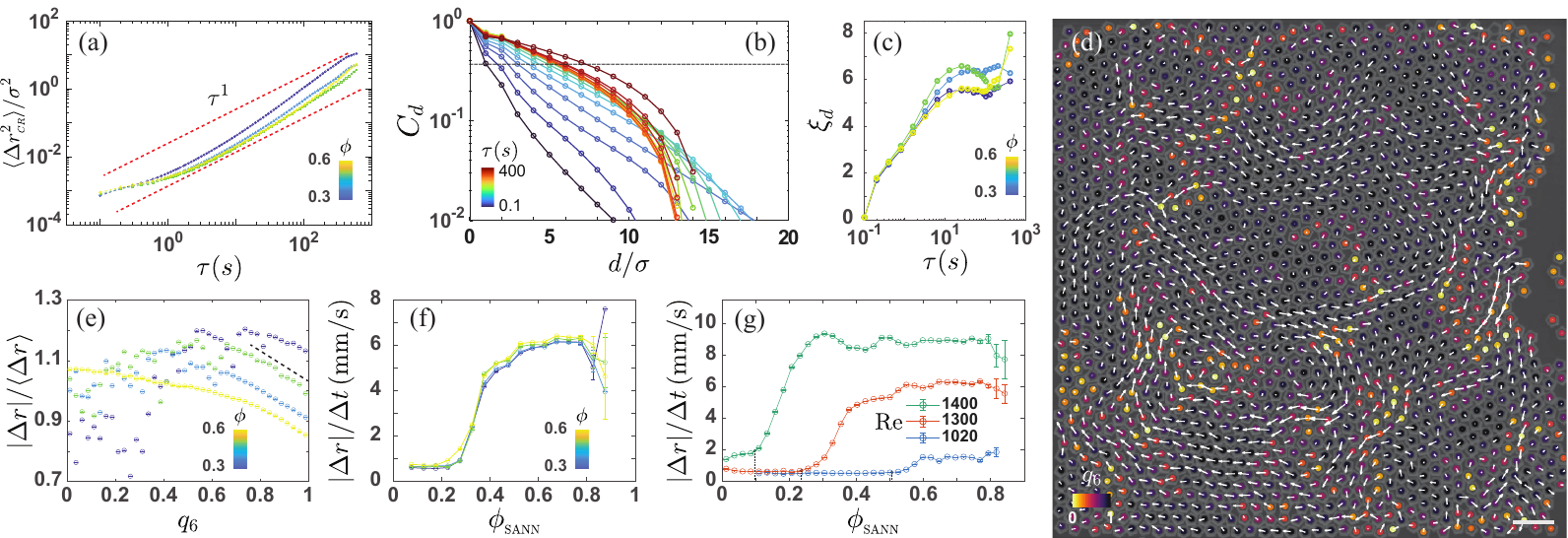}
\caption{(a) Cage-relative mean-squared displacements (MSDs) normalized by $\sigma^2$ at $\mathrm{Re}=1300$ for $\phi=0.3$--$0.6$ (colored by $\phi$), plotted versus lag time $\tau$, where particle displacements are computed relative to their initial Voronoi neighbors. The red dashed line indicates $\tau^1$ scaling. (b) Spatial displacement correlation function $C_d(\tau)$ as a function of distance $d/\sigma$ calculated at different lag times, at $\mathrm{Re}=1300$ and $\phi=0.6$. (c) Correlation length $\xi_d$ extracted from $C_d(\tau)$ as a function of $\tau$ at $\mathrm{Re}=1300$ for $\phi=0.3$--$0.6$. (d) Experimental snapshot at $\mathrm{Re}=1300$ and $\phi=0.6$, overlaid with particle displacement vectors over $\tau=1$\,s and colored by $q_6$. Scale bar: 2\,cm. (e) Normalized particle displacement $|\Delta r|/\langle\Delta r\rangle$ versus $q_6$ at $\mathrm{Re}=1300$ for $\phi=0.3$--$0.6$ (colored by $\phi$), computed over $\tau=1.5$\,s. (f) Mean particle speed $|\Delta r(\Delta t)|/\Delta t$ versus local packing fraction $\phi_{\mathrm{SANN}}$ at $\mathrm{Re}=1300$ for $\phi=0.3$--$0.6$, with $\Delta t=0.1$\,s. (g) Mean particle speed versus $\phi_{\mathrm{SANN}}$ for three air flow conditions. The onset  density increases with decreasing $Re$: $\phi_{\mathrm{SANN}} \approx 0.10$, $0.24$, and $0.51$ for $\mathrm{Re}=1400$, $1300$, and $1020$, respectively.
}
\label{fig:spatialvcorr}
\end{figure*}

We next focus on the super-diffusivity observed in the active-liquid phase, which is a distinct feature that is absent in isolated pentagons as well as the previous sphere packings~\cite{abate2006approach,keys2007measurement}. In this case, the observed super-diffusivity should have a collective, many-body origin related to the hollow pentagon geometry. To understand the origin, we study several cases with $\mathrm{Re}=1300$ and at various $\phi\in[0.3, 0.6]$.

To further confirm the anomalous transport occurs within particle clusters, we calculate the cage-relative MSDs~\cite{illing2017mermin}, see Fig.~\ref{fig:spatialvcorr}a, all showing super-diffusivity at long $\tau$, which rules out rigid-body cluster motion.
Interestingly, at an intermediate $\tau$, we see sub-diffusivity, indicating that particles are caged and tend to keep their neighbors. This suggests that the long-term super-diffusivity involves collective motion of small neighborhoods of particles. 
Indeed, we see large and spatially correlated particle displacements in the experimental snapshot in Fig.~\ref{fig:spatialvcorr}d, where we overlaid the displacement vectors over $\tau=1$\,s (onset of $\alpha>1$).

To further quantify the observed correlation, we calculate the spatial displacement correlation between a particle $i$ and its neighbors $j$ at distance $d=|\vec{r}_i-\vec{r}_j|$,
\begin{equation}
C_d(\tau)=\langle\frac{{\Delta \vec{r}_i}(\tau)\cdot{\Delta \vec{r}_j}(\tau)}{|{\Delta \vec{r}_i}(\tau)|\cdot|{\Delta \vec{r}_j}(\tau)|}\rangle.
\end{equation}
The result in Fig.~\ref{fig:spatialvcorr}b shows that $C_d(\tau)$ decays exponentially, $C_d(\tau)=\exp(-d/\xi_d)$, with the fitted correlation length $\xi_d$ for each $\tau$ shown in Fig.~\ref{fig:spatialvcorr}c. While $\xi_d$ is initially small, reflecting jiggling motion due to random collisions and air-induced stochasticity, it gradually increases with $\tau$, consistent with the observed large correlation in Fig.~\ref{fig:spatialvcorr}d. A plateau is then reached, followed by a slight decrease, which may correspond to long-term diffusion that de-correlates particle motion. A secondary rise is seen in the dense cases, indicating global movement of large clusters. 

In Fig.~\ref{fig:spatialvcorr}d, we further color the particle centers by their $q_6$, which indicates that the large displacements tend to occur around more disordered regions with smaller $q_6$. To quantify this trend, we bin-averaged the displacement of the particles according to their $q_6$ and normalized by the average displacement of all particles, see Fig.~\ref{fig:spatialvcorr}e. Particles with larger displacements generally have smaller $q_6$, suggesting that locally disordered regions are more ``active.''

One plausible explanation for this relation between a particle's activity and its surrounding structure is that when air flows past a pentagon packing structure with low symmetry, a net lateral force can be exerted on the structure as the air pressure in the gaps surrounding this structure cannot be balanced. As pentagons cannot pack with perfect translational and rotational symmetry, all particles experience such unbalanced lateral forces and therefore they are effectively active due to such many-body aerodynamic interactions, especially for particles with low $q_6$. As our particles are monodisperse and pack with hexagonal order at intermediate densities~\cite{schilling2005monte}, the more active regions tend to exist at grain boundaries between ordered domains.

A second important factor for this emergent activity is the aforementioned density-dependence (Fig.~\ref{fig:msd_phase}b), where $\phi$ influences the static-motion threshold. Figure~\ref{fig:spatialvcorr}f  further shows the relation between the particle speed, $|\Delta r(\Delta t)|/\Delta t$ with $\Delta t=0.1$\,s, and the local packing fraction $\phi_{\mathrm{SANN}}$, calculated using the solid-angle nearest neighbor (SANN) method~\cite{vanMeel2012}. Particles in denser local packings tend to move faster, which is opposite to the typical behavior observed in active Brownian particles exhibiting motility-induced phase separation (MIPS)~\cite{Solon2015}. Notably, the data from different global packing fractions $\phi=0.3$--$0.6$ collapse onto a single curve, suggesting that the local packing fraction $\phi_{\mathrm{SANN}}$ is the primary determinant of particle speed in the active-liquid phase. We further measured this relation under three air flow speeds (Fig.~\ref{fig:spatialvcorr}g), all showing the similar density-dependence. At a given $\phi_{\mathrm{SANN}}$, larger $\mathrm{Re}$ results in higher particle speed and smaller onset density (at which the particle speed significantly rises). This is because a densely packed structure forces air to accelerate through its narrow gaps, effectively increasing the local Reynolds number and enhancing stochastic particle motion.

All the evidence above relates the observed super-diffusivity in the active-liquid regime to three important many-body effects from air fluidization, which are the attraction between particles, the increased mobility with lower structural ordering, and the increased velocity fluctuation with increased local density. To summarize: (i) The attraction ensures that particles at intermediate packing fractions can condense like a liquid such that close packing with short-range order exists. (ii) The close-packed pentagon-shaped particles collectively experience lateral forces from the up-lifting air flow if their local structure lacks symmetry. (iii) The increased velocity fluctuations with denser local structures facilitate the activity while preventing MIPS. 

We point out that the gas-like phase, as well as previous air-fluidized experiments with spheres, lacks the particle attraction that brings particles to a close packing, whereas the solid-like phase develops a strong caging effect that dominates the particle dynamics. Thus, both these phases do not exhibit super-diffusivity.

\section{Mimicking the effective many-body activity via a  density-dependent numerical model} \label{sec:sim_anti_active}

To better understand the super-diffusive behavior observed in the active-liquid regime, 
we numerically tailor an {\it ad hoc} confined 2D system of active pentagon-like particles which follow Langevin dynamics, see Materials and Methods. 
To recover the experimental super-diffusivity at intermediate densities, arising from lateral forces induced by a local pressure imbalance due to disorder, we introduce an effective many-body active force acting on each simulated particle. 

\begin{figure*}[t!]
    \centering
 \includegraphics[width=0.9\linewidth]{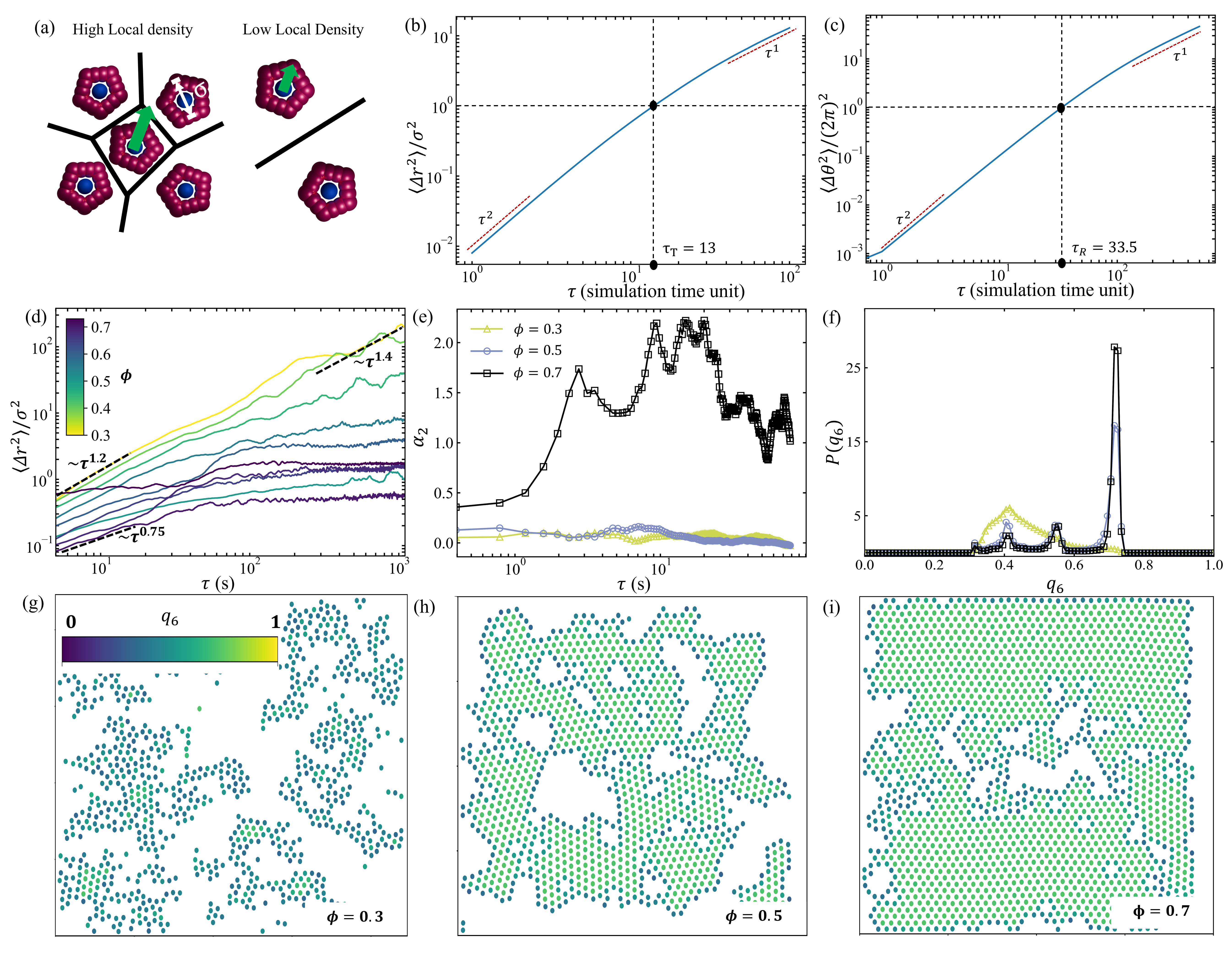}
\vspace{-10 pt}
\caption{(a) Pentagon model. The rigid structure is composed of a blue central particle and $15$ red particles on the sides. The green arrow indicates the direction of the orientation vector. The schematics illustrates the idea of our implemented algorithm shown in section \ref{sec:sim_anti_active}. When the local density of a pentagon is higher, the magnitude of the velocity of the pentagon of concern is also higher; vice versa. 
(b) Mean squared displacement of a single pentagon in an unconfined system. In our model, the activity is positively correlated with the local density; in the case of a super dilute system, our pentagons are supposed to move passively. (c) Mean squared angular displacement of a single pentagon in an unconfined system. (d) Mean squared displacement for the systems with varied $\phi$ ranging from $0.3$ to $0.73$. Dotted lines label the local slope of the curves. (e) Non-Gaussian parameter over time for three different packing fractions $\phi = 0.3, 0.5$ and $0.7$. (f) Probability distribution of $q_6$. (g - i) Snapshots of simulations with $\phi = 0.3$, $0.5$ and $0.7$, with particle centers colored by $q_6$ values. 
}
\label{fig:pentagon_simulation}
\vspace{-5 pt}    
\end{figure*}

The experimentally observed positive relation between a pentagon's local density and speed (Fig.~\ref{fig:spatialvcorr}f,g) motivated us to tailor a numerical model in which the active force, set by the self-propulsion speed, depends on the local density and is aligned with the pentagon's velocity: $\vec{F}_a(\rho)=\gamma_t \vec{v}_a(\rho)$, as sketched in Fig.~\ref{fig:pentagon_simulation}a. Here, $\gamma_t$ is the translational friction coefficient in the Langevin dynamics and $\rho$ is the local density around a single pentagon obtained by Voronoi tessellation ($\rho = A_{voro}^{-1}$, with $A_{voro}$ the area of a Voronoi cell). 

The relationship between speed and local density is modeled by a saturating function,
\begin{equation}
    \vec{v}_a(\rho) = v_0 \left(1 - e^{-\frac{\rho}{\rho_0}}\right)\frac{\vec{v}}{|\vec{v}|}, 
    \label{Eq_vi}
\end{equation}
where $v_0$ is the maximum active speed, $\rho_0$ is a characteristic density scale, and $\rho$ and $\rho_0$ are expressed in units of $1/\text{length}^2$ (as opposed to $\phi$, which is dimensionless). This form captures the experimentally observed saturation of particle speed at 
high local density.

We then simulated a 2D monolayer of confined active pentagons at several packing fractions comparable to the experiments. {\color{black} To map numerical to experimental results, we set the time and the length units to be $0.392$\,s and $1.62$\,mm, respectively (more details in Materials and Methods). In our simulations, $v_0 = 3.53$\,mms$^{-1}$, and $\rho_0=N_{\mathrm{pents}}/A_\mathrm{sys}$}, where $N_\mathrm{pents}$ is the total number of pentagons and $A_\mathrm{sys}$ is the area of the domain. 

Following the modified Langevin dynamics, an isolated pentagon ($\rho\ll\rho_0$) behaves like a Brownian particle, see Fig.~\ref{fig:pentagon_simulation}b,c, in which we report the mean squared displacement and the mean squared angular displacement, respectively.    
Both results show an initial $t^2$ increase and a gradual change to a diffusive behavior ($\alpha=1$).

At intermediate packing fractions, $\phi \in [0.3, 0.5)$, we recover the super-diffusive behavior, as indicated by the MSD's slope $~t^{1.4}$ in Fig.~\ref{fig:pentagon_simulation}d. This originates from the density-dependent activity, which is larger when particles cluster together as in Fig.~\ref{fig:pentagon_simulation}g,h. As the density increases even further, $\phi>0.5$, the system is more crowded and the pentagons' motion more caged. This leads to a sub-diffusive behavior, with the MSD reaching a plateau as soon as  $\phi$ approaches 0.7.

As for microscopic particles' dynamics, Figure~\ref{fig:pentagon_simulation}e shows that $\alpha_2$ takes small values at all time intervals whenever the packing fraction is low/medium ($\phi=0.3$ and 0.5). By contrast, $\alpha_2$  remarkably departs from zero as soon as the packing fraction is $\phi=0.7$ (fingerprint of a non-Gaussian distribution of particle displacement). In terms of the local crystalline structure, the distribution of $q_6$ (Fig.~\ref{fig:pentagon_simulation}f) indicates that for $\phi=0.3$, there is no strong  ordering since a single peak occurs at $q_6\approx0.4$. For $\phi={0.5,0.7}$, we detect three different peaks, corresponding to different local structures, with a higher peak at $q_6=0.75$, representing a higher degree of particles' ordering. 

\section{Discussion}

We experimentally study a monolayer of pentagon particles undergoing  an up-lifting air flow, which causes many-body aerodynamic interactions leading to a collective behavior. Depending on their geometry, whether solid or hollow, pentagons experience effective repulsions or attractions, respectively. In the latter case, hollow pentagons form clusters, in which pentagons display an emergent activity. This is a collective effect that originates when locally disordered structures interact with the up-lifting air flow, resulting in correlated motion across broad temporal and spatial scales. To understand the many-body nature of such an effect, we tailor a {\it coarse grain} numerical model of attractive active pentagons in which activity is an effective result of a locally high number of neighbors.

By contrasting the numerical system of attractive active pentagons with the experimental system of air-fluidized particles  with many-body interactions, we gain further insights from both similarities and differences in the observed dynamics and structures. 
The fact that the experimentally observed super-diffusivity (Fig.~\ref{fig:spatialvcorr}a) can be captured in simulations with local density-dependent self-propelling velocities (Fig.~\ref{fig:pentagon_simulation}d) is consistent with the hypothesis that clustered experimental particles are indeed active and experience additional lateral forces due to the local disorder. 
Note that in both experiments and simulations, the super-diffusivity is facilitated by the mechanisms that particle attractions encourage cluster formation, and that clustered particles experience higher activity (Fig.~\ref{fig:spatialvcorr}f and Eq.~\ref{Eq_vi}). 
In this sense, activity emerges collectively and is coupled to the local disorder, which may give rise to distinct complicated dynamics and structures of the dense clusters.  

While both systems reach a caged and solid-like stage at high packing densities, their behaviors at intermediate packing densities, $\phi\in[0.5,0.7)$, are rather distinct. 
In the numerical system, particles form ``static'' clusters as they are trapped by their neighbors, which could be the combined effect of the attraction and geometrical interlocking.
The structure that pentagons form in numerical simulations (Fig.~\ref{fig:pentagon_simulation}h) is fractal-like, with stable voids dispersed in between. 
By contrast, in the experimental system,  
particles initially form small clusters (Fig.~\ref{fig:app}bi), which later coarsen into a single large cluster without significant voids. Its $q_6$ distribution (Fig.~\ref{fig:msd_phase}f) is smoother and broader than that of simulations (Fig.~\ref{fig:pentagon_simulation}f). More importantly, particles in the experimental clusters are still super-diffusive at the intermediate densities, exhibiting large-scale spatially and temporally correlated motion that is coupled to the structural disorder. 
This is strikingly different from numerical simulations. 

In experiments, the air-mediated particle interaction has a many-body nature, since nearby particles move collectively to minimize their blockage of the air flow. 
Thus, this experimental activity is tightly coupled to attractions. Whereas, in computer simulations we have decoupled the density-dependent activity from attraction, and simulated the former as the result of an effective many-body interaction, while the latter is given by a Lennard-Jones-like, two-body interaction.
The resulting differences in system behaviors, as mentioned above, share similarities with the theoretical results proposed for animal huddles in which each animal would generate heat while moving according to the temperature gradient to optimize its degree of comfort, which was approximated by a three-body interaction~\cite{Zampetaki2021}. 
Such interactions also give rise to aperiodic clustering structures and collective particle reorganization that are distinct from purely two-body interacting systems. 
Our findings from the experimental and numerical systems will facilitate establishing universal features across different complex systems, while inform design principles for intelligent swarms using non-living components.

\section{Materials and Methods}

\noindent \textbf{Experimental details:} A monolayer of pentagon-shaped particles is placed on an air-permeable substrate, as depicted in Fig.~\ref{fig:app}a. Particles have a regular pentagon shape with a height of 3 mm and an altitude (vertex-to-edge distance) of $\sigma=7.2$\,mm. 
To alter the air pressure distribution between particles, we make  particles hollow with a wall thickness of $t_\mathrm{p}=1.2$\,mm. To enhance air-fluidization, we made the inner wall slightly tapered with the bottom being thicker than the top. The particles are 3D-printed using a Stratasys J850 3D PolyJet printer with an accuracy of $20-85$\,$\mu$m.
The air flow is generated by two centrifugal air blowers and channeled through a duct with a circular cross-section and a diameter of $61$\,cm. A polyurethane foam diffuser with a thickness of $19.05$\,mm and porosity of $1.77$\,pore-per-mm is used to regulate the air so that particles experience a uniform and up-lifting incoming air flow. 
The air flow speed can be controlled within the range of $U\in[2.0,3.5]$\,m/s. 

Density-controlled experiments were performed to study emergent collective behaviors. A square arena of width $W=31.1$\,cm was created by pinning carbon fiber rods (diameter $1$\, mm) to the substrate to minimize disturbance to the air flow. The particle area fraction is defined as $\phi=N_\mathrm{pents}A_p(\sigma)/W^2$, where $N_\mathrm{pents}$ is the total number of particles and $A_p$ is the particle area including its hollow center. We explored $\phi\in[0.3,0.73]$, corresponding to $800–2000$ particles, to capture both dilute and dense regimes. Each experiment was initialized by first letting the particles experience high air flow to randomize their positions, after which the air blowers were turned off and restarted at the desired air flow. Particle motion was recorded at $10$\,fps for $12,000$ frames, with the final $6,000$ frames ($10$ min) used for analysis. To characterize single-particle dynamics, a single pentagon was placed on the substrate with the fan already running at the desired air flow and recorded at $100$ fps for $10,000$ frames. In both cases, particles were tracked using a combination of custom MATLAB algorithms and ImageJ TrackMate.

\noindent\textbf{Simulation details:} Since hollow air-fluidized pentagons tend to have attractive interaction with each other, the simulated system consists of a two-dimensional suspension of pentagon-like particles.
The pentagon shape of each particle is approximated by 15 overlapping particles (along the pentagon's edges) and a particle at the geometric center of the pentagon: particles are bound together and move as a rigid body, as shown in Fig.~\ref{fig:pentagon_simulation}a. The resulting pentagon has a fixed radius of $R_p = 2.5\sigma_\textrm{sub}$, where $\sigma_\textrm{sub}$ is the size of each particle. 
The inter-particle attraction seen in experiments is modeled by using the standard Lennard-Jones (L-J) potential to include the interaction between edge particles belonging to different pentagons.
Each pentagon's edges consist of overlapping spheres (with diameter $\sigma_{\mathrm{sub}}$),
interacting with each other via a harmonic potential (to keep the pentagon shape) and an attractive  Lennard-Jones potential, 
\begin{equation}
    U(r)=
    \begin{cases}
        4\epsilon\large[\large(\frac{\sigma_{\mathrm{sub}}}{r}\large)^{12}-\large(\frac{\sigma_{\mathrm{sub}}}{r}\large)^6\large] , & r < r_c \\
        0, & r \geq r_c
    \end{cases}
    \label{Eq_LJ}
\end{equation}
where $r$ is the distance between pairs of spheres within a cutoff distance of    
$r_c = 2.5\sigma_\mathrm{sub}$ so that pentagons could have net attractive interactions over a relatively long range. 
The total external interaction force ($\vec{\nabla} U(\vec{r})$) and torque ($\tau_\mathrm{int}$) acting on each pentagon are obtained by summing the contributions from all its constituent particles.

Pentagons evolve according to Langevin dynamics at constant temperature $T$ [simulated using a modified version of LAMMPS \cite{LAMMPS}], via the following translational (Eq.~\ref{Eq_langevin_Trans}) and rotational (Eq.~\ref{Eq_langevin_Rot}) equations:
\begin{equation}
M \frac{d\vec{v}}{dt} =  \vec{F}_a(\rho)-\vec{\nabla} U(\vec{r}) - \gamma_t \vec{v} + \sqrt{2k_B T\gamma_t}\,\vec{\xi},
\label{Eq_langevin_Trans}
\end{equation}
\begin{equation}
I \frac{d\omega}{dt} = \tau_{\mathrm{int}} - \gamma_r \omega_z + \sqrt{2k_B T\gamma_r}\,\eta.
\label{Eq_langevin_Rot}
\end{equation}

Each pentagon has mass $M$, position $\vec{r}$, velocity $\vec{v}$ and unit orientation vector $\vec{n}$. The parameters $\gamma_t$ and $\gamma_r$ denote the translational and rotational friction coefficients, respectively, and $k_B$ is the Boltzmann constant. The friction coefficients $\gamma_{t,r}$ are related to the diffusion coefficients (translational $D_t$ and rotational $D_r$) via $D_{t,r}=k_BT/\gamma_{t,r}$. The Langevin thermostat is included via $\vec{\xi}$ and $\eta$, which are Gaussian uncorrelated white noise, satisfying $\langle \xi_{i}(t) \xi_{j}(t') \rangle = 2 k_B T \gamma_t \delta_{ij} \delta(t - t')$ and $\langle \eta(t) \eta(t') \rangle = 2 k_B T \gamma_r \delta(t - t')$. 
{\color{black} In our simulations, the energy unit is $k_BT=1$, $\gamma = 1.0$ in simulation reduced units, $D_r = D_t = 1.0$ in simulation reduced units. }

\noindent\textbf{Mapping simulation to experiments:} To match the time and length scale between experiments and simulations, two parameters were considered. For the time scale, $\tau_T$ is used, which is the time when the normalized MSD reaches $1$. Here, $\tau_T$ has a value of $13$ in simulation time units, shown in Fig.~\ref{fig:pentagon_simulation}b, while $\tau_T = 5.1$\,s as indicated in Fig.~\ref{fig:singleP}a. Therefore, we can conclude that 1 simulation time unit is equivalent to $0.392$\,s. 
For the length scale, $\sigma$ is used which labels the altitude of a pentagon (Fig.~\ref{fig:app}). It has a value of $\sigma = 7.2$\,mm in experiments and $\sigma = 4.472$ in simulation length units. 

\begin{acknowledgments}
H.X., Y.F., and W.L. would like to acknowledge the funding from the National Science Foundation grant CMMI-2519512. C.V. acknowledges funding from IHRC22/00002 and Proyecto PID2022-140407NB-C21 by MCIN/AEI/10.13039/501100011033 and FEDER, UE. M.R.-G. and R.F.-Q.G. acknowledge support from Ramón y Cajal program (RYC2021-032055-I) funded by MCIN/AEI/10.13039/501100011033 and by European Union NextGenerationEU/PRTR, a Research Grant from HFSP (Ref.-No: RGEC33/2024) with the award DOI \href{https://doi.org/10.52044/HFSP.RGEC332024.pc.gr.194170} {\nolinkurl{10.52044/HFSP.RGEC332024.pc.gr.194170}} and grant PID2023-147067NB-I00 funded by MCIU/AEI/10.13039/501100011033 and by ERDF/EU. Google Gemini and Google Scholar Lab were used for literature search and Claude Opus 4.8 was used for checking the spelling and grammar of the manuscript.
\end{acknowledgments}

\bibliography{ref}
\newpage

\setcounter{equation}{0}
\setcounter{figure}{0}
\setcounter{table}{0}
\renewcommand{\theequation}{S\arabic{equation}}
\renewcommand{\thefigure}{S\arabic{figure}}
\renewcommand{\thetable}{S\arabic{table}}

\section{Appendixes}
\subsection*{Image processing workflow}
Images were first corrected for lens aberrations to minimize geometric distortion. Basic binarization and background subtraction were applied to enhance particle visibility and reduce noise. Particle segmentation was then performed in MATLAB using a combination of filtering, morphological operations (including opening, closing, erosion, and dilation), and false-positive exclusion based on Voronoi area. This ensures accurate particle identification across a range of area fractions in each experiment.

\begin{figure}[!h]
\centering
\includegraphics[width=0.45\textwidth]{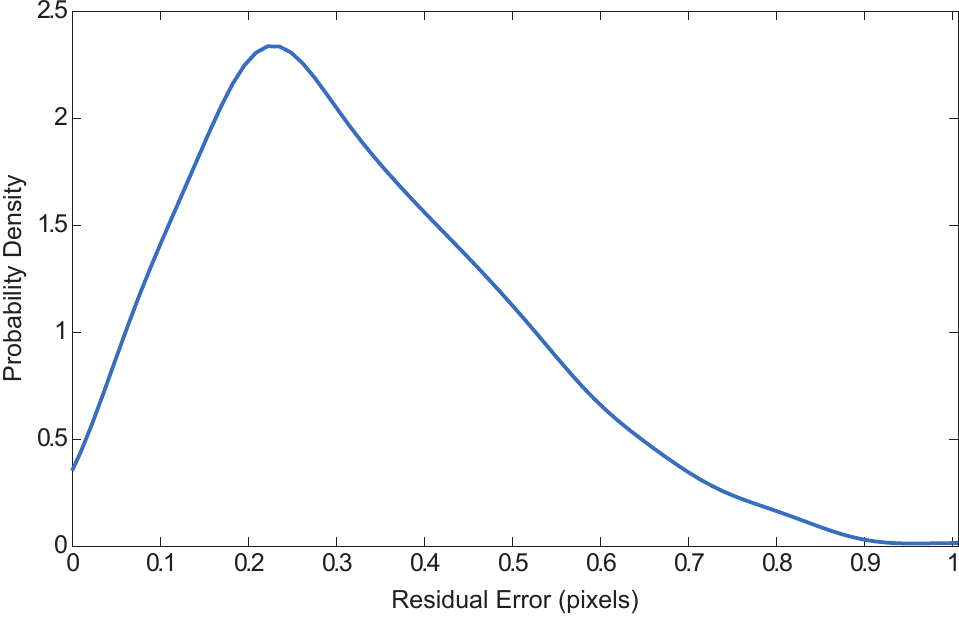}
\caption{\label{fig:tracking_error} 
The probability density function of the error in the TrackMate particle localization result, in comparison with results obtained via manual segmentation. Here, 1\,pixel =\,0.41 mm.
}
\end{figure}

The corrected images were then used for particle tracking via Fiji/ImageJ using the TrackMate plugin with the label detector. To validate the accuracy of particle center positions identified by TrackMate, we compared them to (i) MATLAB regionprops centroids and (ii) manually segmented particles (segmented in MATLAB and measured in ImageJ via Analyze/Measure). A systematic offset of 0.5 pixels between the coordinate systems of MATLAB and TrackMate was corrected. The residual error between manual segmentation and TrackMate was estimated from 500 randomly selected particles and found to have a mean of 0.32 pixels, corresponding to 0.13\,mm, with the probability density of the errors shown in Fig.~\ref{fig:tracking_error}. This is approximately 2\% of the particle size and therefore insignificant in comparison to the particle displacement results we report. 

\subsection*{Estimation of inter-particle forces}
\subsubsection*{Experiments}
\begin{figure*}[htbp]
\centering
\includegraphics[width=\textwidth]{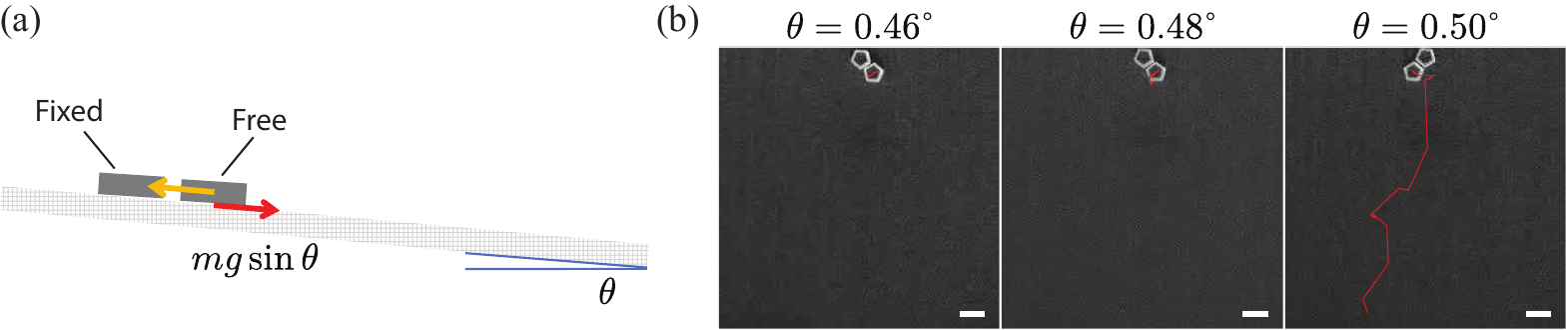}
\caption{\label{fig:tilt} Auxiliary experiments for estimating inter-particle attraction. (a) Schematic of the tilted-substrate setup: one hollow particle is fixed while a second particle is free to drift under a background airflow. The substrate is tilted at an angle $\theta$, and the induced gravitational force component, $mg\sin(\theta)$ is labeled as the red arrow. The orange arrow represents the inter-particle attraction. (b) Representative outcomes as the tilt angle increases: the free particle remains near the fixed particle, drifts locally around it, or moves away; trajectories are shown in red. Scale bar: $1\,$cm.}
\end{figure*}

To quantify the effective attraction between a pair of hollow pentagon particles, we introduced an in-plane body force by tilting the sieve by an angle of $\theta$, as sketched in Fig.~\ref{fig:tilt}a. The background airflow was held constant at $3.2 \,\mathrm{m/s}$. In this configuration, one hollow pentagonal particle was fixed to the substrate, while a second identical hollow particle remained free to move on the tilted surface in the vicinity of the fixed particle. We increased the tilt incrementally until the free particle was just able to overcome the attractive interaction and begin to drift away, as seen in Fig.~\ref{fig:tilt}b; we define this onset as the threshold tilt angle.
Across three trials, the average threshold angle was found to be $\theta=0.48^\circ$. At threshold, the downslope component of the particle’s weight balances the effective attraction, giving $F_{\mathrm{att}} = mg\sin\theta$. Using $m = 5.2\times10^{-5}\,\mathrm{kg}$ and $g=9.8\,\mathrm{m/s^{2}}$, we obtain $\mathrm{F}_\mathrm{att}\approx 4\times10^{-6}\,\mathrm{N}$. 

\subsubsection*{Direct numerical simulations}
\begin{figure*}[b]
\centering
\begin{subfigure}{0.4\textwidth}
    \centering
    \includegraphics[width=\textwidth]{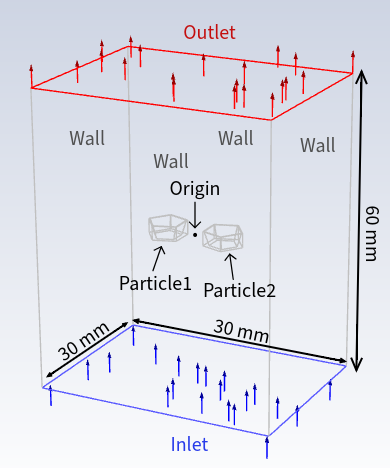}
    \caption{}
    \label{fig:geometry}
\end{subfigure}
\hfill
\begin{subfigure}{0.45\textwidth}
    \centering
    \includegraphics[width=\textwidth]{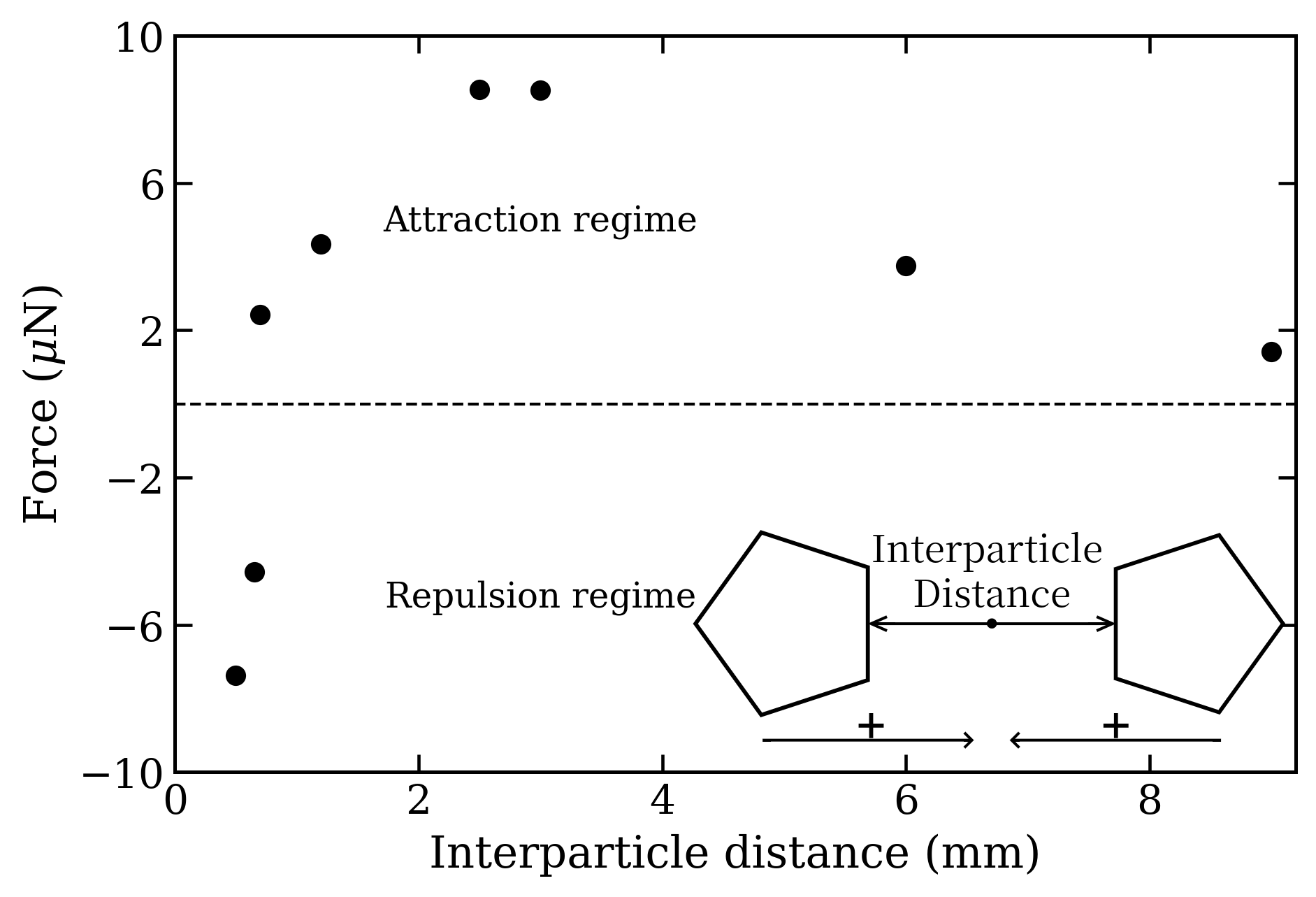}
    \caption{}
    \label{fig:graph}
\end{subfigure}

\vspace{0.5cm}
\caption{Computational fluid dynamics simulation of air flow past a pair of static hollow pentagons. (a) Simulation setup. (b) Effective particle interaction force between two particles at different distances of separation.}
\label{fig:sim_results}
\end{figure*}

To further quantify the interactions between hollow pentagon particles, we performed computational fluid dynamics (CFD) simulations using ANSYS Fluent, which is a finite volume-based CFD solver. For these simulations, the standard k-$\epsilon$ turbulence model was employed to estimate the time-averaged aerodynamic interaction force between two nearby particles; the unsteady wake forces responsible for single-particle stochastic motion are not represented. As shown in Fig.~\ref{fig:sim_results}a, two particles having the same geometry as the experimental particles were fixed in place and simulated at different inter-particle distances. The particles were placed at the center of the computational domain, with pentagon faces oriented toward each other. The domain was chosen to provide sufficient clearance between the particles and walls, so that the boundaries did not strongly influence the flow around the particles. The inlet and outlet were also placed sufficiently far from the particles to allow the flow to develop fully. We simulated an inlet flow with a uniform velocity of 3.0\,m/s. At the outlet, we restricted reverse flow, i.e., downward velocity, to match the experimental scenario. Since the overall air flow corresponds to a low-velocity, incompressible regime, a pressure-based solver was selected. The mesh element size was selected to be smaller than the inter-particle distance to properly resolve the velocity and pressure variations occurring within the gap between the particles. In these simulations, a mesh element size of 0.175\,mm was used, resulting in approximately $2.07 \times 10^{7}$ mesh elements in the domain.

The aerodynamic force acting on each particle was calculated directly from ANSYS Fluent. The force in the dominant interaction axis was considered (the direction along the particle center-center vector). A positive force indicates attraction between the particles, whereas a negative force indicates repulsion. The force was then averaged over the two forces collected from the two particles, which is shown in Fig.~\ref{fig:sim_results}b as a function of the inter-particle distance. At short distances, the inter-particle force is repulsive. As the inter-particle distance increases, the force transitions from repulsive to attractive, reaching a maximum around $9\times10^{-6}\,\mathrm{N}$. This is on the same order of magnitude as the experimental estimation, noting that in the experiments particles do not have exact face-face alignment and the free particle experiences stochastic motion. This attraction force further decays over a length of approximately one particle size ($\sigma=$7.2\,mm).

\end{document}